\documentclass[pra,superscriptaddress,showpacs,eqsecnum,amsmath,twocolumn]{revtex4}
\usepackage{graphicx}
\usepackage{bm}

\begin{document}
%---------- MACROS ----------------------------------------------
\newcommand{\Tr}{\mathop{\mathrm{Tr}}\nolimits}

%---------- FRONT MATTERS ---------------------------------------
\preprint{WU-HEP-04-6}
\title{Distillation of Entanglement between Distant Systems by Repeated Measurements on Entanglement Mediator}
\author{G. Compagno}\email{compagno@fisica.unipa.it}
\author{A. Messina}\email{messina@fisica.unipa.it}
\affiliation{INFM, MURST and Dipartimento di Scienze Fisiche ed Astronomiche dell'Universit\`a di Palermo, Via Archirafi 36, 90123 Palermo, Italy}
\author{Hiromichi Nakazato}\email{hiromici@waseda.jp}
\affiliation{Department of Physics, Waseda University, Tokyo 169-8555, Japan}
\author{A. Napoli}\email{napoli@fisica.unipa.it}
\affiliation{INFM, MURST and Dipartimento di Scienze Fisiche ed Astronomiche dell'Universit\`a di Palermo, Via Archirafi 36, 90123 Palermo, Italy}
\author{Makoto Unoki}
\author{Kazuya Yuasa}\email{yuasa@hep.phys.waseda.ac.jp}
\affiliation{Department of Physics, Waseda University, Tokyo 169-8555, Japan}
\date[]{July 31, 2004}

\begin{abstract}
A recently proposed purification method, in which the Zeno-like measurements of a subsystem can bring about a distillation of another subsystem in interaction with the former, is utilized to yield entangled states between distant systems.
It is shown that the measurements of a two-level system locally interacting with other two spatially separated not coupled subsystems, can distill entangled states from the latter irrespectively of the initial states of the two subsystems.
\end{abstract}
\pacs{03.67.Mn, 03.65.Xp}
\maketitle

\section{Introduction}
One of the key technologies for quantum information and computation is purification/distillation of quantum states \cite{pp,BEZ00}.
Particular pure states, such as entangled states, often play significant roles there, but it is not easy to find such ``clean'' states in nature.
It is therefore required to prepare them out of mixed states; otherwise, we can not carry out any interesting ideas of quantum information and computation.

A new purification mechanism has recently been proposed \cite{NTY03}\@.
It is shown that repeated measurements on a system, say A, result in a purification of another system, say B, in interaction with A \cite{note:QZE}.
That is, the state of B is driven to a pure state irrespectively of its (generally mixed) initial state, if certain conditions are satisfied.
Remarkably, if appropriate adjustment of the relevant parameters is possible, the maximal yield, which is prescribed by the initial mixed state of B and its target pure state, can be attained, while keeping the maximal fidelity, by a finite number of measurements on A (an ``optimal purification").
This constitutes a remarkable contrast to the standard purification protocol \cite{pp,BEZ00}, in which it is generally difficult to realize both a non-vanishing yield and the maximal fidelity at the same time.

Since an entangled state is one of the pure states of two quantum systems, say A and B, one can think of a possibility of extracting the entangled state between A and B by repeatedly performing measurements on X which interacts with both A and B\@.
This possibility has already been pointed out \cite{NUY04,WLS04} and explored to show that one of the Bell states can be extracted when this mechanism is applied to a three-qubit system, where qubits A and B always interact with the other qubit X on which one and the same measurement is repeatedly performed.
Notice that in this case the two systems A and B are not spatially separated, because they are supposed to locally interact with X\@.
On the other hand, it is often required, e.g., in the ideas of quantum teleportation and communication \cite{BEZ00,TeleCom}, to establish an entanglement between two quantum systems that are located at or at least can be sent, without loosing the entanglement, to distant places.
In this respect, it would be worth remembering that interesting ideas of generating an entanglement between two cavities \cite{M02+BP03} and of transferring an entanglement between two modes in a cavity to that between other two modes in different cavities \cite{NMC03} have been proposed.
In the former a two-level atom is sent to interact successively with the two cavities resulting in the generation of an entanglement between the two, and in the latter the entanglement is shown to be transferred by a two-level atom which passes through the two cavities and interacts with the relevant cavity modes.
In this paper, this kind of successive interaction with two quantum systems is incorporated within the framework of the new purification mechanism \cite{NTY03,NUY04,WLS04} to show that an entanglement can be established between the states of the two systems spatially separated (or that can be separated).
Notice that the entangled state is distilled from an arbitrary initial state that is in general mixed, while in the generation of entanglement in Ref.~\cite{M02+BP03} the initial state should be prepared in an appropriate pure state and in the transfer of entanglement in Ref.~\cite{NMC03} the state is assumed to be initially entangled.

After a brief review of the new purification mechanism in Sec.~\ref{sec:rev}, a scheme of successive interaction is introduced in a three-qubit system, A+B+X, in which system X is assumed to interact first with system A and then with B, in Sec.~\ref{sec:qqX}\@. System X is prepared in an initial pure state and is measured after it has interacted with A and B\@.
Then only those events in which system X is found in the initial state are kept. This process will be repeated many times. It is shown that an optimal entanglement purification is actually realizable for a particular choice of interaction and by properly adjusting interaction times and strengths between A and X and B and X\@.
In Sec.~\ref{sec:ccX}, another example of an entanglement purification is examined in a physical system where a two-level atom X is injected successively to the two cavities A and B ``back and forth,'' interacts with their cavity modes under the rotating-wave approximation and the state of X is repeatedly measured in a prescribed way.
It is explicitly shown that, under certain conditions, a particular entangled state between the lowest two modes of each cavity is extracted, irrespectively of the initial cavity states.
Finally we summarize the results obtained and give future perspectives in Sec.~\ref{sec:sum}.

\section{Purification via repeated measurements}
\label{sec:rev}
Let the total system consist of two parts, system A and system B, and the dynamics be described by the total Hamiltonian
\begin{equation}
H=H_\text{A}+H_\text{B}+H_\text{int}, \label{eq:HA+B}
\end{equation}
where $H_\text{int}$ stands for the interaction between the two (sub)systems. We initially prepare the system in a product state
\begin{equation}
\rho_0=|\phi\rangle\langle\phi|\otimes\rho_\text{B}(0)
\label{eq:rho0}
\end{equation}
at $t=0$.
Notice that system B can be in an \textit{arbitrary mixed} state $\rho_\text{B}(0)$.
We perform measurements on A at regular intervals $\tau$ to confirm that it is still in the state $|\phi\rangle$ \cite{note:QZE}, while the total system A+B during the time $\tau$ evolves unitarily in terms of the total Hamiltonian $H$.
Since the measurement is performed only on system A, the action of such a (projective, for simplicity) measurement can be conveniently described by the following projection operator
\begin{equation}
\mathcal{O}\equiv|\phi\rangle\langle\phi|\otimes\hat{1}_\text{B}.
\label{eq:calO}
\end{equation}
Thus the state of system A is set back to $|\phi\rangle$ every after $\tau$, while that of B just evolves dynamically on the basis of the total Hamiltonian $H$. We repeat the same measurement, represented by (\ref{eq:calO}), $N$ times and collect only those events in which system A has been found in the state $|\phi\rangle$ consecutively $N$ times; other events are discarded. The state of system B is then described by the density matrix
\begin{equation}
\rho_\text{B}^{(\tau)}(N)
=\bigl(V_\phi(\tau)\bigr)^N\rho_\text{B}(0)
 \bigl(V_\phi^\dagger(\tau)\bigr)^N/P^{(\tau)}(N),
\label{eq:rhoBN}
\end{equation}
where
\begin{equation}
V_\phi(\tau)\equiv\langle\phi|e^{-iH\tau}|\phi\rangle
\label{eq:Vphi}
\end{equation}
is an operator acting on B and
\begin{align}
P^{(\tau)}(N)
&=\Tr\Bigl[(\mathcal{O}e^{-iH\tau}\mathcal{O})^N\rho_0
 (\mathcal{O}e^{iH\tau}\mathcal{O})^N\Bigr]\nonumber\\
&=\Tr_\text{B}\Bigl[\bigl(V_\phi(\tau)\bigr)^N\rho_\text{B}(0)
  \bigl(V_\phi^\dagger(\tau)\bigr)^N\Bigr]
\label{eq:PtauN}
\end{align}
is the probability for these events to occur (yield).
This normalization factor appearing in (\ref{eq:rhoBN}) reflects the fact that only the right outcomes are collected in this process.

In order to examine the asymptotic state of system B for large $N$, consider, assuming its existence, the spectral decomposition of the operator $V_\phi(\tau)$, which is not hermitian, $V_\phi(\tau)\not=V_\phi^\dagger(\tau)$. We therefore need to set up both the right- and left-eigenvalue problems
\begin{equation}
V_\phi(\tau)|u_n\rangle=\lambda_n|u_n\rangle,\quad \langle
v_n|V_\phi(\tau)=\lambda_n\langle v_n|. \label{eq:unvn}
\end{equation}
The eigenvalue $\lambda_n$ is complex valued in general, but its absolute value is bounded \cite{NUY04}
\begin{equation}
\quad0\leq|\lambda_n|\leq1. \label{eq:bnded}
\end{equation}
This reflects the unitarity of the time evolution operator $e^{-iH\tau}$\@.
These eigenvectors are assumed to form a complete orthonormal set in the following sense
\begin{equation}
\sum_n|u_n\rangle\langle v_n|=\hat1_\text{B},\quad \langle
v_n|u_m\rangle=\delta_{nm}. \label{eq:conset}
\end{equation}
(We normalize $|u_n\rangle$ as $\langle u_n|u_n\rangle=1$, while the norm of $\langle v_n|$ has been fixed by the above relations and is not necessarily unity.)
The operator $V_\phi(\tau)$ itself is now expanded in terms of these eigenvectors
\begin{equation}
V_\phi(\tau)=\sum_n\lambda_n|u_n\rangle\langle v_n|.
\label{eq:Vphiuv}
\end{equation}
It is now easy to see that the $N$th power of this operator is expressed as
\begin{equation}
\bigl(V_\phi(\tau)\bigr)^N=\sum_n\lambda_n^N|u_n\rangle\langle
v_n| \label{eq:VphiN}
\end{equation}
and therefore it is dominated by a single term for large $N$
\begin{equation}
\bigl(V_\phi(\tau)\bigr)^N
\xrightarrow{\text{large }N} \lambda_0^N|u_0\rangle\langle v_0|,
\label{eq:Vinf}
\end{equation}
provided the largest (in magnitude) eigenvalue $\lambda_0$ is \textit{discrete, nondegenerate and unique}.
If these conditions are satisfied, the density operator of system B is driven to a pure state
\begin{equation}
\rho_\text{B}^{(\tau)}(N)
\xrightarrow{\text{large }N}
|u_0\rangle\langle u_0|
\label{eq:rhoBinf}
\end{equation}
with the probability
\begin{equation}
P^{(\tau)}(N)
\xrightarrow{\text{large }N}
|\lambda_0|^{2N}
\langle v_0|\rho_\text{B}(0)|v_0\rangle. \label{eq:Pinf}
\end{equation}
The pure state $|u_0\rangle$, which is nothing but the right-eigenvector of the operator $V_\phi(\tau)$ belonging to the largest (in magnitude) eigenvalue $\lambda_0$, is thus distilled in system B\@.
This is the purification scheme proposed in \cite{NTY03}.

A few comments are in order.
First, the final pure state $|u_0\rangle$ toward which system B is to be driven is dependent on the choice of the state $|\phi\rangle$ on which system A is projected every after measurement, on the measurement interval $\tau$ and the Hamiltonian $H$, but does not depend on the initial state of system B at all.
In this sense, the purification is accomplished irrespectively of the initial (mixed) state $\rho_\text{B}(0)$.
Second, as is clear in the above exposition, what is crucial in this purification scheme is the repetition of one and the same measurement (more appropriately, spectral decomposition) and the measurement interval $\tau$ need not be very small \cite{note:QZE}.
It instead remains an adjustable parameter.
Third, if we can make other eigenvalues than $\lambda_0$ much smaller in magnitude
\begin{equation}
|\lambda_n/\lambda_0|\ll1\quad\text{for } n\not=0,
\label{eq:quick}
\end{equation}
by adjusting parameters, we will need fewer steps (i.e., smaller $N$) to purify system B\@.

It is now evident that the purification can be made \textit{optimal}, if the conditions (\ref{eq:quick}) and
\begin{equation}
|\lambda_0|=1 \label{eq:noloss}
\end{equation}
are satisfied.
This condition (\ref{eq:noloss}) assures that we can repeat as many measurements as we wish without running the risk of losing the yield $P^{(\tau)}(N)$ in order to make the fidelity to the target state $|u_0\rangle$,
\begin{equation}
F^{(\tau)}(N)\equiv\Tr_\text{B}
\left[\rho_\text{B}^{(\tau)}(N)|u_0\rangle\langle u_0|
\right], \label{eq:F}
\end{equation}
higher. Actually, the yield $P^{(\tau)}(N)$ decays like
\begin{align}
P^{(\tau)}(N)
&=\sum_{n,m}\lambda_n^N\lambda_m^*{}^N\langle
v_n|\rho_\text{B}(0)|v_m\rangle \langle u_m|u_n\rangle\nonumber\\
&\xrightarrow{\text{large }N}
|\lambda_0|^{2N}
\langle v_0|\rho_\text{B}(0)|v_0\rangle \label{eq:Pasymp}
\end{align}
and the condition (\ref{eq:noloss}) can bring us with the non-vanishing yield $\langle v_0|\rho_\text{B}(0)|v_0\rangle$ even in the $N\to\infty$ limit.
Therefore the condition (\ref{eq:noloss}) makes the two (sometimes not compatible) demands, i.e., \textit{higher fidelity and non-vanishing yield}, achievable, with fewer steps when the condition (\ref{eq:quick}) is met. In this sense, the purification is considered to be optimal.

It would be desirable if an optimal purification can be realized by an appropriate choice of the state $|\phi\rangle$ and/or tuning of the measurement interval $\tau$ and parameters in a given Hamiltonian.
A few simple systems have already been examined \cite{NTY03,NUY04,WLS04} to show how such optimal purifications are made possible.

\section{Entanglement distillation in a two-qubit system A+B by an entanglement mediator X} \label{sec:qqX}
Since an entangled state of system A+B is one of the pure states, there is a possibility that we apply the purification mechanism described above to distill the initial (generally mixed) state of A+B to a desired entangled state.
This possibility has already been pointed out and it has been explicitly demonstrated that one of the Bell states $|\Psi^-\rangle$ of the two qubit systems A+B can actually be extracted if we repeatedly measure one and the same state of another qubit system X, the interactions of which are symmetrical with respect to\ A and B, resulting with the maximal yield \cite{NUY04,WLS04}\@.
One of the limitations of this model is that the entanglement can be established only when the two systems A and B locally interact with the same system X at the same time and therefore it does not seem to allow to establish an entanglement between two systems that are spatially separated. A new scheme is certainly needed.

When the two systems A and B are spatially separated, their local interactions with the other system X can not take place simultaneously.
This means that the interactions are considered to become effective one by one, i.e., system X first interacts with, say system A and then with B \cite{M02+BP03,NMC03}\@.
This kind of process can be conveniently described by a time-dependent total Hamiltonian. In this section, a total system composed of three qubits (or three spin $1/2$ systems), A, B, and X, is considered.
The two qubits A and B interact with the other qubit X successively, the state of X, initially prepared in a particular state, say in up state, is measured after the interactions with A and B, and only those events in which the state of X is found in up state will be retained and other events are just discarded.
The process will be repeated many times and we are interested in the resulting state of A+B\@.

Assume that the three-qubit system A+B+X is described by a time-dependent total Hamiltonian $H(t)$.
Qubit X, which is initially prepared in up state $|{\uparrow}_\text{X}\rangle$, is first brought to interaction with qubit A for time interval $t_\text{A}$.
The Hamiltonian in this period is
\begin{subequations}
\begin{equation}
H(t)=H_0+H_\text{XA}'. \label{eq:HXA}
\end{equation}
Next, after a free time evolution under the free Hamiltonian $H_0$ for time duration $\tau_\text{A}$, qubit X interacts with another qubit B, which has no direct interaction with A, for $t_\text{B}$.
The Hamiltonian for this period reads
\begin{equation}
H(t)=H_0+H_\text{XB}'. \label{eq:HXB}
\end{equation}
\end{subequations}
After another free time evolution for $\tau_\text{B}$, the state of X is measured and only those cases in which qubit X is found in its initial up state $|{\uparrow}_\text{X}\rangle$ shall be retained.
The whole process, i.e.,
\begin{align}
&\text{interaction between X and A for $t_\text{A}$}\nonumber\\
&\quad\to\text{free evolution for $\tau_\text{A}$}\nonumber\\
&\quad\to\text{interaction between X and B for $t_\text{B}$}\nonumber\\
&\quad\to\text{free evolution for $\tau_\text{B}$}\nonumber\\
&\quad\to\text{projection to\ }|{\uparrow}_\text{X}\rangle
\label{eq:wp}
\end{align}
shall be repeated many times and we are interested in the final state of A+B\@. See Fig.~1.
\begin{figure}
\includegraphics[width=0.48\textwidth]{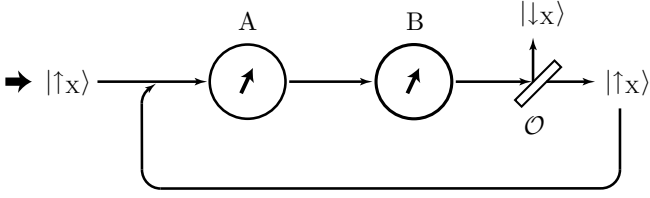}
\caption{Qubit X, prepared in $|{\uparrow}_\text{X}\rangle$, is brought to interaction with qubits A and B successively and then its state is measured.  If it is found in $|{\uparrow}_\text{X}\rangle$, the whole process is repeated again; other events are discarded.}
\label{fig:1}
\end{figure}

In order to describe the above process explicitly and to show the possibility of entanglement distillation by this process, we consider the following free and interaction Hamiltonians
\begin{subequations}
\begin{gather}
H_0=\frac{\omega}{2}(1+\sigma_3^\text{(A)})+\frac{\omega}{2}(1+\sigma_3^\text{(B)})
    +\frac{\omega}{2}(1+\sigma_3^\text{(X)}),
\label{eq:H0}\\
H_\text{XA}'=g_\text{A}\sigma_1^\text{(X)}\sigma_1^\text{(A)},\quad
H_\text{XB}'=g_\text{B}\sigma_1^\text{(X)}\sigma_1^\text{(B)},
\label{eq:HAB}
\end{gather}
\end{subequations}
where $\sigma_i^\text{(A)}$'s are Pauli matrices acting on the Hilbert space of system A and so on, and $g_\text{A}$ and $g_\text{B}$ are real (assumed, for definiteness, to be positive) coupling constants.
The free Hamiltonians for A, B, and X are assumed to be the same for simplicity and are characterized solely by the common energy gap $\omega$.

It is clear that the relevant evolution operator for the whole process (\ref{eq:wp}) is given by
\begin{multline}
V\equiv\langle{\uparrow}_\text{X}|e^{-iH_0\tau_\text{B}}e^{-i(H_0+H_\text{XB}')t_\text{B}}\\
{}\times e^{-iH_0\tau_\text{A}}e^{-i(H_0+H_\text{XA}')t_\text{A}}|{\uparrow}_\text{X}\rangle.
\label{eq:V}
\end{multline}
It is an elementary task to evaluate this operator, since each factor on the right-hand side is easily evaluated in terms of the eigenstates of the Hamiltonian in the exponent. Indeed, we have
\begin{widetext}
\begin{subequations}
\begin{multline}
e^{-i(H_0+H_\text{XA}')t_\text{A}}|{\uparrow}_\text{X}\rangle
=e^{-i\omega t_\text{A}(1+\sigma_3^\text{(B)})/2-i\omega
t_\text{A}}
  \Bigl[|{\uparrow}_\text{X}\rangle
  \{(\cos\varphi_\text{A}-i\sin\varphi_\text{A}\cos2\theta_\text{A})
  |{\uparrow}_\text{A}\rangle\langle{\uparrow}_\text{A}|+\cos(
  g_\text{A}t_\text{A})
  |{\downarrow}_\text{A}\rangle\langle{\downarrow}_\text{A}|\}\\
  {}+|{\downarrow}_\text{X}\rangle
  \{-i\sin\varphi_\text{A}\sin2\theta_\text{A}
  |{\downarrow}_\text{A}\rangle\langle{\uparrow}_\text{A}|-i\sin
  (g_\text{A}t_\text{A})
  |{\uparrow}_\text{A}\rangle\langle{\downarrow}_\text{A}|\}\Bigr],
\label{eq:rf}
\end{multline}
and similarly
\begin{multline}
\langle{\uparrow}_\text{X}|e^{-i(H_0+H_\text{XB}')t_\text{B}}
=e^{-i\omega t_\text{B}(1+\sigma_3^\text{(A)})/2-i\omega
t_\text{B}}
  \Bigl[
  \langle{\uparrow}_\text{X}|
  \{(\cos\varphi_\text{B}-i\sin\varphi_\text{B}\cos2\theta_\text{B})
  |{\uparrow}_\text{B}\rangle\langle{\uparrow}_\text{B}|+\cos
  (g_\text{B}t_\text{B})
  |{\downarrow}_\text{B}\rangle\langle{\downarrow}_\text{B}|\}
\displaybreak[0]\\
  {}+\langle{\downarrow}_\text{X}|
  \{-i\sin\varphi_\text{B}\sin2\theta_\text{B}
  |{\uparrow}_\text{B}\rangle\langle{\downarrow}_\text{B}|-i\sin
  (g_\text{B}t_\text{B})
  |{\downarrow}_\text{B}\rangle\langle{\uparrow}_\text{B}|\}
  \Bigr],
\label{eq:lf}
\end{multline}
\end{subequations}
\end{widetext}
where the angles $\varphi_\text{A(B)}$ and $\theta_\text{A(B)}$ are defined as
\begin{subequations}
\label{eq:angles}
\begin{gather}
\varphi_\text{A(B)}=t_\text{A(B)}\sqrt{\omega^2+g_\text{A(B)}^2},\\
\sin2\theta_\text{A(B)}=\frac{g_\text{A(B)}}{\sqrt{\omega^2+g_\text{A(B)}^2}},\\
\cos2\theta_\text{A(B)}=\frac{\omega}{\sqrt{\omega^2+g_\text{A(B)}^2}}.
\end{gather}
\end{subequations}

Let us introduce a parity operator $\mathcal{P}\equiv\sigma_3^\text{(A)}\sigma_3^\text{(B)}$ whose eigenvalues $+1$ and $-1$ single out two subspaces of the product Hilbert space $\mathcal{H}_\text{A}\otimes H_\text{B}$ invariant under the action of the operator $V$. The two states $|{\uparrow}_\text{A}{\uparrow}_\text{B}\rangle$ and $|{\downarrow}_\text{A}{\downarrow}_\text{B}\rangle$ generate the even parity subspace and the following $2\times2$ matrix $\mathcal{M}$ with its elements
\begin{subequations}
\begin{align}
\mathcal{M}_{11}
={}&e^{-i\omega(t_\text{A}+2\tau_\text{A}+t_\text{B}+2\tau_\text{B})}
 (\cos\varphi_\text{A}-i\sin\varphi_\text{A}\cos2\theta_\text{A})
\nonumber\\
 &{}\times(\cos\varphi_\text{B}-i\sin\varphi_\text{B}\cos2\theta_\text{B}),
\label{eq:M11}\\
\mathcal{M}_{12} ={}&-e^{-i\omega
t_\text{A}}\sin\varphi_\text{A}\sin2\theta_\text{A} \sin
(g_\text{B}t_\text{B}),
\label{eq:M12}\\
\mathcal{M}_{21} ={}&-e^{-i\omega(t_\text{B}+2\tau_\text{B})}
 \sin (g_\text{A}t_\text{A})\sin\varphi_\text{B}\sin2\theta_\text{B},
\label{eq:M21}\displaybreak[0]\\
\mathcal{M}_{22} ={}&\cos (g_\text{A}t_\text{A})\cos
(g_\text{B}t_\text{B}) \label{eq:M22}
\end{align}
\end{subequations}
allows to completely characterize the action of $V$ in this subspace as follows
\begin{equation}
V\left[\begin{matrix}
       |{\uparrow}_\text{A}{\uparrow}_\text{B}\rangle\\
\noalign{\smallskip}
          |{\downarrow}_\text{A}{\downarrow}_\text{B}\rangle
       \end{matrix}\right]
=e^{-i\omega(t_\text{A}+\tau_\text{A}+t_\text{B}+\tau_\text{B})}{\mathcal{M}} \left[\begin{matrix}
      |{\uparrow}_\text{A}{\uparrow}_\text{B}\rangle\\
\noalign{\smallskip}
      |{\downarrow}_\text{A}{\downarrow}_\text{B}\rangle
      \end{matrix}\right].
\label{eq:p+}
\end{equation}
We proceed in the same way for the odd parity subspace spanned by the states $|{\uparrow}_\text{A}{\downarrow}_\text{B}\rangle$ and $|{\downarrow}_\text{A}{\uparrow}_\text{B}\rangle$.
To this end we define the $2\times2$ matrix $\mathcal{N}$ with its elements
\begin{subequations}
\begin{align}
&\mathcal{N}_{11} =e^{-i\omega(2\tau_\text{A}+t_\text{B})}
 (\cos\varphi_\text{A}-i\sin\varphi_\text{A}\cos2\theta_\text{A})
 \cos (g_\text{B}t_\text{B}),
\label{eq:N11}\\
&\mathcal{N}_{12}
=-\sin\varphi_\text{A}\sin2\theta_\text{A}\sin\varphi_\text{B}\sin2\theta_\text{B},
\label{eq:N12}\\
&\mathcal{N}_{21}
=-e^{-i\omega(t_\text{A}+2\tau_\text{A}+t_\text{B})}\sin
(g_\text{A}t_\text{A})\sin (g_\text{B}t_\text{B}),
\label{eq:N21}\\
&\mathcal{N}_{22} =e^{-i\omega(t_\text{A}+2\tau_\text{A})}
 \cos (g_\text{A}t_\text{A})(\cos\varphi_\text{B}
 -i\sin\varphi_\text{B}\cos2\theta_\text{B}),
\label{eq:N22}
\end{align}
\end{subequations}
so that the action of $V$ is represented as
\begin{equation}
V\left[\begin{matrix}
      |{\uparrow}_\text{A}{\downarrow}_\text{B}\rangle\\
\noalign{\smallskip}
      |{\downarrow}_\text{A}{\uparrow}_\text{B}\rangle
      \end{matrix}\right]
=e^{-i\omega(t_\text{A}+t_\text{B}+2\tau_\text{B})}\mathcal{N}
\left[\begin{matrix}
      |{\uparrow}_\text{A}{\downarrow}_\text{B}\rangle\\
\noalign{\smallskip}
      |{\downarrow}_\text{A}{\uparrow}_\text{B}\rangle
      \end{matrix}\right].
\label{eq:p-}
\end{equation}

In order to show explicitly that the process (\ref{eq:wp}) with the particular choice of interaction (\ref{eq:HAB}) admits an entanglement distillation for qubit-system A+B, it turns out to be enough to consider a much simpler case.
Let us treat systems A and B symmetrically, except for the ordering of their interactions with system X\@.
We choose the same parameters for A and B, i.e., $g_\text{A}=g_\text{B}\equiv g$, $t_\text{A}=t_\text{B}\equiv t$ and $\tau_\text{A}=\tau_\text{B}\equiv\tau$ ($\varphi_\text{A(B)}\to\varphi$ and $\theta_\text{A(B)}\to\theta$).
For the parity-odd states, the matrix $\mathcal{N}$ now is simplified to be
\begin{widetext}
\begin{equation}
\mathcal{N} =\left(\begin{matrix}
       e^{-i\omega(t+2\tau)}\cos (gt)(\cos\varphi-i\sin\varphi\cos2\theta)
       &-\sin^2\varphi\sin^22\theta\\
       \noalign{\medskip}
       -e^{-2i\omega(t+\tau)}\sin^2(gt)
       &e^{-i\omega(t+2\tau)}\cos (gt)(\cos\varphi-i\sin\varphi\cos2\theta)
       \end{matrix}\right).
\label{eq:N}
\end{equation}
\end{widetext}
It is easy to find the condition under which an entangled state of the form
\begin{equation}
|\Psi\rangle\equiv\frac{1}{\sqrt2}(|{\uparrow}_\text{A}{\downarrow}_\text{B}\rangle+e^{i\chi}
|{\downarrow}_\text{A}{\uparrow}_\text{B}\rangle), \label{eq:Psi}
\end{equation}
where $\chi$ is a real parameter, is an eigenstate of this matrix
$\mathcal{N}$ (and therefore, of the operator $V$).
A straightforward calculation shows that if the parameters $g$, $t$ and $\tau$ are so chosen that the following relation
\begin{equation}
\cos\varphi-i\sin\varphi\cos2\theta=-e^{i\omega\tau}\cos (gt),
\label{eq:opt1}
\end{equation}
is satisfied, the state $|\Psi\rangle$ with $\chi=\omega(t+\tau)$ is indeed an eigenstate of $V$ belonging to the eigenvalue $\lambda_0=-e^{-3i\omega(t+\tau)}$.
[There is another possibility of optimal distillation of the above entangled state (\ref{eq:Psi}), but with a different $\chi$, i.e., $\chi=\omega(t+\tau)+\pi$.
This case is realized under the condition (\ref{eq:opt1}) with the replacement $\omega\tau\to\omega\tau+\pi$; the corresponding eigenvalue is also given by the shifted one, i.e., $e^{-3i\omega(t+\tau)}$.]

Notice that  we are not allowed to set $\cos(gt)\sin(gt)=0$ because it would result in a degenerate (in magnitude) eigenvalue.
Observe that we have essentially two conditions (\ref{eq:opt1}), while we have three independent combination of parameters $gt$, $\omega t$ and $\omega\tau$.
We, therefore, have a possibility of an optimal distillation of the entangled state $|\Psi\rangle$, if the magnitudes of the other eigenvalues of $V$ are made smaller than unity.
The remaining eigenvalue of $\mathcal{N}$ under the conditions in (\ref{eq:opt1}) reads
\begin{multline}
e^{-i\omega(t+\tau)}\Bigl[e^{-i\omega\tau}\cos
(gt)(\cos\varphi-i\sin\varphi\cos2\theta)+\sin^2(gt)\Bigr]\\
=e^{-i\omega(t+\tau)}
[-\cos^2(gt)+\sin^2(gt)], \label{eq:ev'}
\end{multline}
and its absolute value can not be made unity when $\cos (gt)\sin(gt)\not=0$.
On the other hand, matrix $\mathcal{M}$ is expressed as
\begin{equation}
\mathcal{M}=\begin{pmatrix}
         e^{-2i\omega(t+\tau)}\cos^2(gt)&\mp e^{-i\omega t}\sin^2(gt)\\
         \noalign{\medskip}
         \mp e^{-i\omega(t+2\tau)}\sin^2(gt)&\cos^2(gt)
         \end{pmatrix}
\label{eq:M1}
\end{equation}
and the absolute values of the eigenvalues of this matrix can not reach unity if $\cos\omega(t+\tau)\not=\pm1$ and $\cos (gt)\sin(gt)\not=0$.
This means that, under these conditions on $gt$, $\omega t$ and $\omega\tau$ satisfying the relation (\ref{eq:opt1}), an optimal purification (i.e., distillation) of the entangled state $(|{\uparrow}_\text{A}{\downarrow}_\text{B}\rangle+e^{i\omega(t+\tau)}|{\downarrow}_\text{A}{\uparrow}_\text{B}\rangle)/\sqrt2$ is possible.
It is in fact easily shown that the left-eigenstate of $V$ belonging to the eigenvalue $\lambda_0=-e^{-3i\omega(t+\tau)}$ is expressed as
\begin{equation}
\langle\Phi|=\frac{1}{\sqrt2}
(\langle{\uparrow}_\text{A}{\downarrow}_\text{B}|+{e^{-i\chi}}\langle{\downarrow}_\text{A}{\uparrow}_\text{B}|),
\quad\langle\Phi|\Psi\rangle=1, \label{eq:Phi}
\end{equation}
and therefore the yield $P^{(\tau)}(N)$ approaches asymptotically, as $N$ becomes large, a finite value
\begin{multline}
P^{(\tau)}(\infty)
=\frac{1}{2}\Bigl[{}\langle{\uparrow}_\text{A}{\downarrow}_\text{B}|+
 {e^{-i\chi}}\langle{\downarrow}_\text{A}{\uparrow}_\text{B}|\Bigr]\\
 {}\times\rho_\text{AB}(0)
 \Bigl[|{\uparrow}_\text{A}{\downarrow}_\text{B}\rangle+e^{i\chi}|{\downarrow}_\text{A}{\uparrow}_\text{B}\rangle\Bigr].
\label{eq:yield}
\end{multline}
This is nothing but the probability of finding the target entangled state $|\Psi\rangle=(|{\uparrow}_\text{A}{\downarrow}_\text{B}\rangle+e^{i\chi}|{\downarrow}_\text{A}{\uparrow}_\text{B}\rangle)/\sqrt2$ in the initial state $\rho_\text{AB}(0)$.

\section{Entanglement distillation of cavity modes}
\label{sec:ccX}
In the previous section, the possibility of realizing an entanglement distillation is demonstrated for the three-qubit system A+B+X\@.
The particular form of interaction (\ref{eq:HAB}) is shown to be suitable for this purpose following the procedure (\ref{eq:wp}).
In this section, another application of the purification mechanism \cite{NTY03,NUY04,WLS04} is explored in a system composed of a two-level system (e.g., an atom) interacting with two single-mode cavities.
The two cavities may be located at spatially distant places (or may be near and separated later) and we aim at extracting an entanglement between the two-cavity states by repeatedly bringing the two-level atom into interaction with them and then selecting a particular state of the atom by measurements.

The ideas of generating \cite{M02+BP03} and of transferring \cite{NMC03} entanglement in two-cavity system have already been proposed and studied in the context of the cavity quantum electrodynamics (CQED) \cite{Science}, spectacularly developed over the last two decades both in the microwave \cite{microwave} and optical \cite{optical} domains.
Entanglement is generated from a properly prepared pure state in the former \cite{M02+BP03} and an initially prepared entanglement in one cavity is transformed into another entanglement between the two cavities in the latter \cite{NMC03}, via successive interactions with a two-level atom.
The atom plays the role of a ``mediator'' or ``transformer'' of entanglement. A similar, but more complicated role is sought for the atom in the present scheme, because its interactions with the cavities and the measurements of its state are expected to enable us to extract an entangled state, that is to produce an entanglement distillation, \textit{irrespectively of the initial states of the two cavities}.

For simplicity, suppose that the two cavities, A and B, are identical and their interaction with a two-level atom X is well described by the Jaynes--Cummings Hamiltonian \cite{JC51}.
Let $a$ and $b$ indicate the annihilation operators of the modes of the two cavities A and B, respectively.
The free and interaction Hamiltonians are
\begin{subequations}
\begin{gather}
H_0=\frac{\omega}{2}(1+\sigma_3)+\omega a^\dagger a+\omega b^\dagger b,\label{eq:H0ab}\\
H'_\text{XA}=g_\text{A}(\sigma_+a+\sigma_-a^\dagger),\quad
H'_\text{XB}=g_\text{B}(\sigma_+b+\sigma_-b^\dagger).
\label{eq:Hab}
\end{gather}
\end{subequations}
A state where the atom X is in up(down) state and the modes $a$ and $b$ are in the $n$th and $m$th levels, respectively, is denoted as $|{\uparrow}({\downarrow}),n,m\rangle$ ($n,\,m=0,1,2,\ldots$).

Since our purpose is to extract a pure state not in product form but entangled, it turns out that simple processes like (\ref{eq:wp}) for the three-qubit system would not work.
Indeed, because of the choice of the Jaynes--Cummings (rotating-wave) interactions (\ref{eq:Hab}), the number operator $(1+\sigma_3)/2+a^\dagger a+b^\dagger b$ commutes with any of $H_0$, $H'_\text{XA}$, and $H'_\text{XB}$, and therefore any state of the two-cavity system of the form $|n,0\rangle$, which is a product state, is easily seen to be an eigenstate of the time evolution operator constructed analogously to (\ref{eq:V}).
(If the down state of X is measured, product states $|0,m\rangle$ are found to be eigenstates of the relevant time evolution operator.)\@
Thus a process different from (\ref{eq:wp}) would be necessary for our purpose.

\begin{figure}[t]
\includegraphics[width=0.48\textwidth]{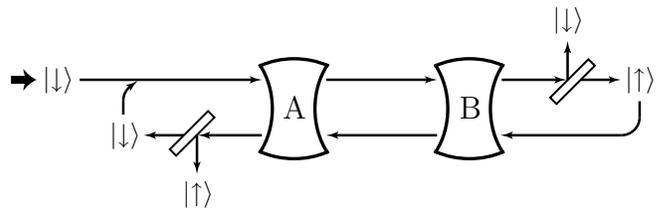}
\caption{A two-level atom X, prepared in $|{\downarrow}\rangle$, is brought to interaction with cavity modes $a$ and $b$ in the two cavities A and B successively, and its state is measured after the interactions. Atoms that are found in state $|{\uparrow}\rangle$ will be sent back to the cavities in the reversed order. The state of atom X is again measured, and if it is found in $|{\downarrow}\rangle$, the whole process is repeated; other events are discarded.}
\label{fig:2}
\end{figure}

The above consideration would suggest that, with interaction given by (\ref{eq:Hab}), it would be better to select after measurement a state of X different from the initial state.
However, at the same time, we need a procedure that can be repeated many times within the present framework of the purification mechanism.
Thus we choose a procedure that can be described schematically as
\begin{align}
&\text{preparation in $|{\downarrow}\rangle$}\nonumber\\
&\quad\to\text{interaction between X and A for $t_\text{A}$}\nonumber\displaybreak[0]\\
&\quad\to\text{free evolution for $\tau_\text{A}$}\nonumber\\
&\quad\to\text{interaction between X and B for $t_\text{B}$}\nonumber\\
&\quad\to\text{free evolution for $\tau_\text{B}$}\nonumber\\
&\quad\to\text{projection to\ }|{\uparrow}\rangle\nonumber\\
&\quad\to\text{free evolution for $\tau_\text{B}$}\nonumber\\
&\quad\to\text{interaction between X and B for $t_\text{B}$}\nonumber\\
&\quad\to\text{free evolution for $\tau_\text{A}$}\nonumber\\
&\quad\to\text{interaction between X and A for $t_\text{A}$}\nonumber\\
&\quad\to\text{projection to\ }|{\downarrow}\rangle.
\label{eq:wp2}
\end{align}
See also Fig.~\ref{fig:2}.

This is clearly a generalization of the purification process, ``projection'' $\to$ ``time evolution'' $\to$ ``projection.''
Indeed, in the above scheme, ``time evolution" is not meant in the usual sense, that is described by a total Hamiltonian.
It is instead interrupted by another projection.
However, the condition under which the purification mechanism does work is essentially the same as in the ordinary cases and all what we have to do here is to investigate the relevant evolution operator corresponding to the above process (\ref{eq:wp2}).
It would be important to notice that the above choice of the initial and projected states for system X is not arbitrary.
In fact if it were prepared in the up state, the procedure analogous to that described before does not work.
The vacuum state of the two cavities, which is a product state, would indeed turn out to be an eigenstate.

The relevant evolution for the above process (\ref{eq:wp2}) is represented by products of the time-evolution and projection operators and each of them is easily evaluated.
The only non-trivial operators are
\begin{widetext}
\begin{subequations}
\begin{multline}
e^{-i(H_0+H'_\text{XA})t_\text{A}} =e^{-i\omega b^\dagger
bt_\text{A}}
 \sum_{n=0}^\infty e^{-i(n+1)\omega t_\text{A}}
       \Bigl(\cos\varphi_\text{A}^{(n+1)}|{\uparrow},n\rangle\langle{\uparrow},n|
             -i\sin\varphi_\text{A}^{(n+1)}|{\uparrow},n\rangle\langle{\downarrow},n+1|\\
       {}-i\sin\varphi_\text{A}^{(n+1)}|{\downarrow},n+1\rangle\langle{\uparrow},n|
       +e^{i\omega t_\text{A}}\cos\varphi_\text{A}^{(n)}
        |{\downarrow},n\rangle\langle{\downarrow},n|
       \Bigr)
\label{eq:Uta}
\end{multline}
and
\begin{multline}
e^{-i(H_0+H'_\text{XB})t_\text{B}} =e^{-i\omega a^\dagger
at_\text{B}}
 \sum_{m=0}^\infty e^{-i(m+1)\omega t_\text{B}}
       \Bigl(\cos\varphi_\text{B}^{(m+1)}|{\uparrow},m\rangle\langle{\uparrow},m|
             -i\sin\varphi_\text{B}^{(m+1)}|{\uparrow},m\rangle\langle{\downarrow},m+1|\displaybreak[0]\\
       {}-i\sin\varphi_\text{B}^{(m+1)}|{\downarrow},m+1\rangle\langle{\uparrow},m|
       +e^{i\omega t_\text{B}}\cos\varphi_\text{B}^{(m)}
       |{\downarrow},m\rangle\langle{\downarrow},m|
       \Bigr),
\label{eq:Utb}
\end{multline}
\end{subequations}
where angles $\varphi_\text{A}^{(n)}$ and $\varphi_\text{B}^{(m)}$ are defined as
\begin{equation}
\varphi_\text{A}^{(n)}\equiv g_\text{A}t_\text{A}\sqrt n,\qquad
\varphi_\text{B}^{(m)}\equiv g_\text{B}t_\text{B}\sqrt m.
\label{eq:vphi}
\end{equation}
It is an elementary task to evaluate in this case the relevant evolution operator $V_\text{c}$, analogously to
\eqref{eq:Vphi},
\begin{equation}
V_\text{c} \equiv
\langle{\downarrow}|e^{-i(H_0+H'_\text{XA})t_\text{A}}e^{-iH_0\tau_\text{A}}
e^{-i(H_0+H'_\text{XB})t_\text{B}}e^{-iH_0\tau_\text{B}}
|{\uparrow}\rangle
\langle{\uparrow}|
e^{-iH_0\tau_\text{B}}e^{-i(H_0+H'_\text{XB})t_\text{B}}
e^{-iH_0\tau_\text{A}}e^{-i(H_0+H'_\text{XA})t_\text{A}}|{\downarrow}\rangle,
\label{eq:Vc}
\end{equation}
and its explicit expression reads
\begin{align}
V_\text{c} =-\sum_{n,m=0}^\infty e^{-2i(n+m)\omega T}
 \Bigl[
 &(\sin^2\varphi_\text{A}^{(n)}\cos^2\varphi_\text{B}^{(m+1)}
 +\cos^2\varphi_\text{A}^{(n)}\sin^2\varphi_\text{B}^{(m)})
 |n,m\rangle\langle n,m|
\nonumber\\
 &{}+\sin\varphi_\text{A}^{(n+1)}\cos\varphi_\text{A}^{(n)}
   \sin\varphi_\text{B}^{(m)}\cos\varphi_\text{B}^{(m)}
   |n+1,m-1\rangle\langle n,m|\nonumber\\
\noalign{\medskip}
 &{}+\sin\varphi_\text{A}^{(n)}\cos\varphi_\text{A}^{(n-1)}
   \sin\varphi_\text{B}^{(m+1)}\cos\varphi_\text{B}^{(m+1)}
   |n-1,m+1\rangle\langle n,m|
   \Bigr],
\label{eq:Vcexp}
\end{align}
\end{widetext}
where $T\equiv t_\text{A}+\tau_\text{A}+t_\text{B}+\tau_\text{B}$.

It is manifest from (\ref{eq:Vcexp}) that in the product Hilbert space of the two cavities there are sectors of $V_\text{c}$ within which the action of $V_\text{c}$ is closed.
These invariant sectors are characterized by the number $n+m$.
We have $(n+m+1)$ states $\{|n+m,0\rangle,\ldots,|0,n+m\rangle\}$ for the ($n+m$)th sector.
Notice that the singlet state (vacuum state) $|0,0\rangle$ ($n+m=0$) belongs to zero-eigenvalue of $V_\text{c}$ and we need not consider it, for $V_\text{c}|0,0\rangle=0$.
This is closely related to the choice of the initial state of X and is the reason why we must prepare X in the down state $|{\downarrow}\rangle$.

Let us turn our attention first to the doublet subspace ($n+m=1$).
The action of $V_\text{c}$ on this subspace is easily read from
(\ref{eq:Vcexp}) as
\begin{widetext}
\begin{equation}
V_\text{c}\left(
         \begin{matrix}
         |1,0\rangle\\
\noalign{\medskip}
         |0,1\rangle
         \end{matrix}
         \right)
=-e^{-2i\omega T}
 \left(
 \begin{matrix}
 \sin^2\varphi_\text{A}^{(1)}\cos^2\varphi_\text{B}^{(1)}
 &\sin\varphi_\text{A}^{(1)}\sin\varphi_\text{B}^{(1)}\cos\varphi_\text{B}^{(1)}\\
\noalign{\smallskip}
 \sin\varphi_\text{A}^{(1)}\sin\varphi_\text{B}^{(1)}\cos\varphi_\text{B}^{(1)}
 &\sin^2\varphi_\text{B}^{(1)}
 \end{matrix}
 \right)\left(
        \begin{matrix}
        |1,0\rangle\\
\noalign{\medskip}
        |0,1\rangle
        \end{matrix}
        \right).
\label{eq:dblt}
\end{equation}
\end{widetext}
Observe that the determinant of this matrix always vanishes, which means that one of the eigenvalues is zero and the other is given by the trace of the matrix $-e^{-2i\omega T}(\sin^2\varphi_\text{A}^{(1)}\cos^2\varphi_\text{B}^{(1)}+\sin^2\varphi_\text{B}^{(1)})$.
Therefore we have a possibility of obtaining the largest (in magnitude) eigenvalue by adjusting the parameter $g_\text{A}t_\text{A}$ so that
\begin{equation}
\sin\varphi_\text{A}^{(1)}\equiv\sin (g_\text{A}t_\text{A})=\pm1.
\label{eq:ccond}
\end{equation}
(The possibility $\cos\varphi_\text{B}^{(1)}=0$ would result, not in an entanglement distillation, but in a product-state purification.)
In such a case, the above eigenvalue equation is simplified to
\begin{align}
V_\text{c}\left(
         \begin{matrix}
         |1,0\rangle\\
\noalign{\medskip}
         |0,1\rangle
         \end{matrix}
         \right)
={}&\left(
 \begin{matrix}
 \cos\varphi_\text{B}^{(1)}&\mp\sin\varphi_\text{B}^{(1)}\\
 \noalign{\smallskip}
 \pm\sin\varphi_\text{B}^{(1)}&\cos\varphi_\text{B}^{(1)}
 \end{matrix}
 \right)
 \left(
 \begin{matrix}
 -e^{-2i\omega T}&0\\
 \noalign{\medskip}
 0&0
 \end{matrix}
 \right)\nonumber\displaybreak[0]\\
&{}\times\left(
 \begin{matrix}
 \cos\varphi_\text{B}^{(1)}&\pm\sin\varphi_\text{B}^{(1)}\\
\noalign{\smallskip}
 \mp\sin\varphi_\text{B}^{(1)}&\cos\varphi_\text{B}^{(1)}
 \end{matrix}
 \right)
 \left(
 \begin{matrix}
 |1,0\rangle\\
 \noalign{\medskip}
 |0,1\rangle
 \end{matrix}
 \right),
\label{eq:dblt2}
\end{align}
from which it is clear that the entangled state $|\Psi_\text{c}^{(1)}\rangle=\cos\varphi_\text{B}^{(1)}|1,0\rangle\pm\sin\varphi_\text{B}^{(1)}|0,1\rangle$ can be extracted with the maximal probability by this setup.
Notice that it still remains the freedom to adjust the value of $\varphi_\text{B}^{(1)}=g_\text{B}t_\text{B}$.

The remaining task is to check whether there are other eigenstates of $V_\text{c}$ belonging to eigenvalues with unit magnitude, under the condition (\ref{eq:ccond}).
Consider the invariant sector characterized by $k=n+m>1$ that is composed of $k+1$ states $\{|k,0\rangle,\ldots,|0,k\rangle\}$.
The action of $V_\text{c}$ on this sector is represented by the following matrix [see (\ref{eq:Vcexp})]
\begin{widetext}
\begin{equation}
V_\text{c} \left(
\begin{matrix}
|k,0\rangle\\
|k-1,1\rangle\\
\vdots\\
|2,k-2\rangle\\
|1,k-1\rangle\\
|0,k\rangle
\end{matrix}
\right) =-e^{-2ik\omega T}
 \left(
  \begin{matrix}
  c_k   &d_k    &\ldots&0     &0     &0\\
  d_k   &c_{k-1}&\ldots&0     &0     &0\\
  \vdots&\vdots &\ddots&\vdots&\vdots&\vdots\\
  0     &0      &\ldots&c_2   &d_2   &0\\
  0     &0      &\ldots&d_2   &c_1   &d_1\\
  0     &0      &\ldots&0     &d_1   &c_0
  \end{matrix}
 \right)
\left(
\begin{matrix}
|k,0\rangle\\
|k-1,1\rangle\\
\vdots\\
|2,k-2\rangle\\
|1,k-1\rangle\\
|0,k\rangle
\end{matrix}
\right), \label{eq:kth}
\end{equation}
\end{widetext}
where matrix elements $c_j,\,d_j$ can be read from (\ref{eq:Vcexp}) as
\begin{align}
c_j&=\sin^2\varphi_\text{A}^{(j)}\cos^2\varphi_\text{B}^{(k-j+1)}
   +\cos^2\varphi_\text{A}^{(j)}\sin^2\varphi_\text{B}^{(k-j)},
\label{eq:cj}\\
\noalign{\smallskip}
d_j&=\sin\varphi_\text{A}^{(j)}\cos\varphi_\text{A}^{(j-1)}
   \sin\varphi_\text{B}^{(k-j+1)}\cos\varphi_\text{B}^{(k-j+1)}.
\label{eq:dj}
\end{align}
It is important to notice that the condition (\ref{eq:ccond}) implies that the element $d_2=\sin\varphi_\text{A}^{(2)}\cos\varphi_\text{A}^{(1)}\sin\varphi_\text{B}^{(k-1)}\times\cos\varphi_\text{B}^{(k-1)}$ vanishes, irrespectively of $\varphi_\text{B}^{(k-1)}$ and thus the sector further splits into two subsectors $\{|k,0\rangle,\ldots,|2,k-2\rangle\}$ and $\{|1,k-1\rangle,\,|0,k\rangle\}$.
Furthermore, it is easily seen that the entangled state in the latter subspace of the form $|\Psi_\text{c}^{(k)}\rangle=\cos\varphi_\text{B}^{(k)}|1,k-1\rangle\pm\sin\varphi_\text{B}^{(k)}|0,k\rangle$ has the eigenvalue $-e^{-2ik\omega T}$, while, as shown in the Appendix, no eigenstate in the former subspace $\{|k,0\rangle,\ldots,|2,k-2\rangle\}$ belongs to a unit (in magnitude) eigenvalue (if $k$ is smaller than 9).

We have seen that there are, for any $k$-sector, many entangled states $|\Psi_\text{c}^{(k)}\rangle=\cos\varphi_\text{B}^{(k)}|1,k-1\rangle\pm\sin\varphi_\text{B}^{(k)}|0,k\rangle$ ($k=1,2,\ldots$) (that increase, in general, in higher $k$-sectors) extracted with the optimal probabilities by the process (\ref{eq:wp2}).
Repeated interactions of the two-level atom X in the cavities A and B and the prescribed measurements (projections) certainly bring us with a statistical (classical) mixture of these entangled states.
The situation would not be considered completely satisfactory, since we would not be able to distill a single entangled state by the process (\ref{eq:wp2}).
There is, however, a way out of this difficulty.
We may prepare such an initial (mixed) state of A+B that contains only those sectors with relatively small $k$s.
Such a preparation of the initial state would effectively eliminate the possibility of obtaining other states than $|\Psi_\text{c}^{(k)}\rangle$ after performing the process (\ref{eq:wp2}).
For example, we may consider the following preparation procedure, which is nothing but a purification process applied to cavity B\@.
We send a two-level atom prepared in down state $|{\downarrow}\rangle$ to cavity B\@.
After its interaction, which is again assumed to be of the form (\ref{eq:Hab}), with B, the atom is measured and only those events in which it is found in the state $|{\downarrow}\rangle$ are retained.
This process is to be repeated many times and the resulting state of A+B, which will be used as the initial state for the following entanglement distillation process (\ref{eq:wp2}), would be dominated by the state $\rho_\text{AB}(0)\sim\rho_\text{A}(0)\otimes|0\rangle\langle0|$, since the vacuum state of system B is the unique eigenstate of the relevant evolution operator $\sim\langle{\downarrow}|e^{-iH'_\text{XB}t}|{\downarrow}\rangle$ belonging to eigenvalue unity if no fine tunings are made on the parameters.
After having prepared the state $\rho_\text{AB}(0)$, we repeat the process (\ref{eq:wp2}) under the condition (\ref{eq:ccond}).
We would finally end up with the single entangled state $|\Psi_\text{c}^{(1)}\rangle$, because our initial state $\rho_\text{AB}(0)$ satisfies $\langle\Phi_\text{c}^{(k)}|\rho_\text{AB}(0)|\Phi_\text{c}^{(k)}\rangle=\langle\Psi_\text{c}^{(k)}|\rho_\text{AB}(0)|\Psi_\text{c}^{(k)}\rangle=0$ for $k>1$, where $\langle\Phi_\text{c}^{(k)}|$ is the left-eigenstate corresponding to $|\Psi_\text{c}^{(k)}\rangle$.

Concluding this section we wish to give typical values of relevant parameters under the aspect of the possibility of implementing our proposal in laboratory.
We concentrate on the estimation of the total duration $T$ of the experiment.
To this end, we choose to be in the context of microwave CQED where both the geometrical arrangement of the experimental set up and the intensity of the atom-field coupling regime, seem more favorable to our proposal.
Let us first note that the typical atom-field coupling constant $g$ ($g_\text{A}$ or $g_\text{B}$) can be chosen in such a way that $g\sim10^4$--$10^5\,\text{s}^{-1}$ \cite{WaltherQ}.
Moreover the lifetime of a Rydberg atom is ${}\gtrsim10^{-2}\,\text{s}$ \cite{Haroche}.
As for the quality factor $Q$ of the cavities currently used in laboratory, we quote typical values of the order of $10^{8}$--$10^{10}$ \cite{WaltherQ,Haroche} corresponding to a cavity damping time $1\,\text{ms}$--$1\,\text{s}$.
Considering that in our case $t_\text{A}\sim t_\text{B}\sim g^{-1}$, the total duration $T$ of the experiment may be estimated and turns out to be compatible with the entanglement distillation  proposed in this section.

\section{Summary}
\label{sec:sum}
In this paper, the idea of extracting entangled states among systems located at spatially separated places, irrespectively of their initial states, has been proposed and applied to simple systems to show the potentiality of a novel measurement-based purification scheme \cite{NTY03}.
The establishment of entanglement distillation relies on the successive interactions between the systems under consideration and the so-called ``mediator" quantum system.
In the first example, it is demonstrated that the entanglement between the two qubit states is possible via their interactions with another qubit, which plays the role of the entanglement mediator, with the maximal yield (optimal distillation).
In the case of distillation of cavity-mode entanglement, however, a modification of the original simple scheme, that is, ``interaction $\to$ measurement $\to$ interaction $\to\cdots$,'' is required and the modified procedure (\ref{eq:wp2}) turns out to result in the entangled state $|\Psi_\text{c}^{(1)}\rangle$, after an appropriate preparation of the initial state.
We stress that there would also be a possibility of obtaining an entangled state not only in the lowest sector $k=1$ but also in the higher $k$ sector, provided an appropriate initial state be prepared.

It would be worth stressing that in spite of such a modification required in the second example, the underlying notion is still the same: the action of a measurement (represented by a projection operator, for simplicity) causes an essential and critical dynamical change, not only in the system measured, but also in the others interacting with the former.
Since the notion is so general, one can devise various applications of this scheme in many different situations.
The examples explored in this paper are just two of them and further applications will be reported elsewhere.

Finally we add some comments on the practical setup of our proposal.
In both schemes reported in the previous sections, the entanglement mediator is an atom appropriately prepared before entering in interaction with the two subsystems to be entangled and subjected to a conditional measurement of its internal state at the end of the two successive coupling processes.
We wish to stress that to realize in practice the required ``many crosses'' scheme one may synchronize the injection of the $j$th atom of the sequence into the process with the successful outcome of the internal state measurement of the $(j-1)$th one.
This kind of experimental setup might be preferable to the conceptually simpler one based on the idea of using always the same atom reversing its direction of motion at exit to re-inject it into the process.
Thus the representations in Figs.~1 and 2, as far as the aspect under scrutiny is concerned, have been reported only for the sake of simplicity.

\begin{acknowledgments}
The authors (H.N., M.U., and K.Y.) acknowledge useful and helpful discussions with Prof.~I. Ohba.
H.N. is grateful for the warm hospitality at Universit\`a di Palermo.
This work is partly supported by a Grant for The 21st Century COE Program (Physics of Self-\hspace*{0mm}Organization Systems) at Waseda University and a Grant-in-Aid for Priority Areas Research (B) (No.~13135221), both from the Ministry of Education, Culture, Sports, Science and Technology, Japan, by a Grant-in-Aid for Scientific Research (C) (No.~14540280) from the Japan Society for the Promotion of Science, by a Waseda University Grant for Special Research Projects (No.~2002A-567) and by the bilateral Italian-Japanese project 15C1 on ``Quantum Information and Computation'' of the Italian Ministry for Foreign Affairs.
\end{acknowledgments}

\appendix*
\section{}
In this Appendix, a symmetric matrix of the form ($k\ge2$)
\begin{equation}
 \left(
 \begin{matrix}
 c_k   &d_k    &0      &\ldots&0     &0\\
 d_k   &c_{k-1}&d_{k-1}&\ldots&0     &0\\
 0     &d_{k-1}&c_{k-2}&\ldots&0     &0\\
 \vdots&\vdots &\vdots &\ddots&\vdots&\vdots\\
 0     &0      &0      &\ldots&c_3   &d_3\\
 0     &0      &0      &\ldots&d_3   &c_2
 \end{matrix}
 \right),
\label{A:matrix}
\end{equation}
with the matrix elements [see (\ref{eq:cj}) and (\ref{eq:dj})]
\begin{subequations}
\label{A:cjdj}
\begin{align}
c_j&=\sin^2\varphi_\text{A}^{(j)}\cos^2\varphi_\text{B}^{(k-j+1)}
   +\cos^2\varphi_\text{A}^{(j)}\sin^2\varphi_\text{B}^{(k-j)},
\label{A:cj}\\
\noalign{\smallskip}
d_j&=\sin\varphi_\text{A}^{(j)}\cos\varphi_\text{A}^{(j-1)}
   \sin\varphi_\text{B}^{(k-j+1)}\cos\varphi_\text{B}^{(k-j+1)},
\label{A:dj}
\end{align}
\end{subequations}
is investigated with particular attention on its eigenstates belonging to the eigenvalues with unit magnitudes.
Since this matrix is real and symmetric, its eigenvalues are all real, and the eigenvalues of relevance in the framework of our procedure here are $\pm1$.

For the first possibility $+1$, let us consider the determinant $I_i$ ($k\ge i\ge2$) defined by
\begin{equation}
I_i\equiv\left|
  \begin{matrix}
  c_i-1 &d_i      &0        &\ldots&0     &0\\
  d_i   &c_{i-1}-1&d_{i-1}  &\ldots&0     &0\\
  0     &d_{i-1}  &c_{i-2}-1&\ldots&0     &0\\
  \vdots&\vdots   &\vdots   &\ddots&\vdots&\vdots\\
  0     &0        &0        &\ldots&c_3-1 &d_3\\
  0     &0        &0        &\ldots&d_3   &c_2-1
  \end{matrix}
  \right|.
\label{A:Ii}
\end{equation}
It is easy to see that the particular form of $I_i$ and the definitions of $c_j$ and $d_j$ in (\ref{A:cjdj}) lead to a recursion relation
\begin{align}
&I_i+\cos^2\varphi_\text{A}^{(i)}\cos^2\varphi_\text{B}^{(k-i)}I_{i-1}\nonumber\\
&\quad=-\sin^2\varphi_\text{A}^{(i)}\sin^2\varphi_\text{B}^{(k-i+1)}\nonumber\\
&\quad\phantom{{}={}}{}\times\left(
 I_{i-1}+\cos^2\varphi_\text{A}^{(i-1)}\cos^2\varphi_\text{B}^{(k-i+1)}I_{i-2}
 \right)
\label{A:rr}
\end{align}
for $k\ge i\ge4$.
This is further reduced to
\begin{multline}
I_i=-\cos^2\varphi_\text{A}^{(i)}\cos^2\varphi_\text{B}^{(k-i)}I_{i-1}\\
{}+(-1)^{i-1}\prod_{j=1}^{i-1}\sin^2\varphi_\text{A}^{(j+1)}\sin^2\varphi_\text{B}^{(k-j)}\\
(k\ge i\ge2). \label{A:rr1}
\end{multline}
If we set $I_i\equiv(-1)^{i+1}P_i$, then $P_i$ is found to be positive semi-definite and to satisfy
\begin{equation}
P_i=\cos^2\varphi_\text{A}^{(i)}\cos^2\varphi_\text{B}^{(k-i)}P_{i-1}
    +\prod_{j=1}^{i-1}\sin^2\varphi_\text{A}^{(j+1)}\sin^2\varphi_\text{B}^{(k-j)}.
\label{A:Pi}
\end{equation}
This relation is easily solved to yield the explicit form of $I_k=(-1)^{k+1}P_k$ with
\begin{widetext}
\begin{align}
P_k
&=\cos^2\varphi_\text{A}^{(2)}\cdots\cos^2\varphi_\text{A}^{(k)}
  \cos^2\varphi_\text{B}^{(1)}\cdots\cos^2\varphi_\text{B}^{(k-2)}
  \nonumber\\
&\quad
 {}+\sum_{n=2}^{k-1}
  \sin^2\varphi_\text{A}^{(2)}\cdots\sin^2\varphi_\text{A}^{(n)}
  \cos^2\varphi_\text{A}^{(n+1)}\cdots\cos^2\varphi_\text{A}^{(k)}
  \cos^2\varphi_\text{B}^{(1)}\cdots\cos^2\varphi_\text{B}^{(k-n-1)}
  \sin^2\varphi_\text{B}^{(k-n+1)}\cdots\sin^2\varphi_\text{B}^{(k-1)}
  \nonumber\\
&\quad
 {}+\sin^2\varphi_\text{A}^{(2)}\cdots\sin^2\varphi_\text{A}^{(k)}
  \sin^2\varphi_\text{B}^{(1)}\cdots\sin^2\varphi_\text{B}^{(k-1)}.
\label{A:Pk}
\end{align}
\end{widetext}

Since each term in (\ref{A:Pk}) has the same sign to each other, vanishing of $I_k$, which is nothing but the condition for the matrix (\ref{A:matrix}) to possess eigenstates that belong to eigenvalue unity, is equivalent to that of each term.
This means that there are $k$ conditions for three parameters $g_\text{A} t_\text{A}$, $g_\text{B} t_\text{B}$ and $k$ and it seems impossible to have a vanishing $I_k$ in general, unless most of the conditions are simultaneously satisfied.
If we choose a particular value for $g_\text{A} t_\text{A}$, say $g_\text{A} t_\text{A}=\pi/2$ as in (\ref{eq:ccond}), however, all $I_k$ with $k\ge9$ vanish because each term in (\ref{A:Pk}) contains either $\sin^2\varphi_\text{A}^{(4)}$ or $\cos^2\varphi_\text{A}^{(9)}$, each of which vanishes.

On the other hand, the matrix (\ref{A:matrix}) is shown to have no eigenstate belonging to the eigenvalue $-1$.
Consider the following determinant
\begin{equation}
J_i\equiv \left|
\begin{matrix}
c_i+1 &d_i      &0        &\ldots&0     &0\\
d_i   &c_{i-1}+1&d_{i-1}  &\ldots&0     &0\\
0     &d_{i-1}  &c_{i-2}+1&\ldots&0     &0\\
\vdots&\vdots   &\vdots   &\ddots&\vdots&\vdots\\
0     &0        &0        &\ldots&c_3+1 &d_3\\
0     &0        &0        &\ldots&d_3   &c_2+1
\end{matrix}
\right|, \label{A:Ji}
\end{equation}
where $c_j$ and $d_j$ are again given in (\ref{A:cjdj}).
Since $c_j>0,\,d_j^2<1$ and therefore $J_2=c_2+1>0$ and $J_3=(c_3+1)(c_2+1)-d_3^2>0$, let us assume that all $J_\ell$'s are positive definite for $\ell$ up to $i-1$.
Then it follows that
\begin{align}
J_i
={}&(c_i+1)J_{i-1}-d_i^2J_{i-2}\nonumber\displaybreak[0]\\
>{}&c_iJ_{i-1}-d_i^2J_{i-2}\nonumber\displaybreak[0]\\
={}&\sin^2\varphi_\text{A}^{(i)}\cos^2\varphi_\text{B}^{(k-i+1)}\nonumber\\
 &{}\times
 \Bigl(
 J_{i-1}-\cos^2\varphi_\text{A}^{(i-1)}\sin^2\varphi_\text{B}^{(k-i+1)}J_{i-2}
 \Bigr)\nonumber\\
&{}+\cos^2\varphi_\text{A}^{(i)}\sin^2\varphi_\text{B}^{(k-i)}J_{i-1}.
\label{A:irr}
\end{align}
This relation recursively yields inequalities
\begin{align}
&J_i-\cos^2\varphi_\text{A}^{(i)}\sin^2\varphi_\text{B}^{(k-i)}J_{i-1}\nonumber\\
&\quad>\sin^2\varphi_\text{A}^{(i)}\cos^2\varphi_\text{B}^{(k-i+1)}\nonumber\\
&\quad\phantom{{}>{}}
{}\times
  \Bigl(
  J_{i-1}-\cos^2\varphi_\text{A}^{(i-1)}\sin^2\varphi_\text{B}^{(k-i+1)}J_{i-2}
  \Bigr)
\nonumber\\
&\quad>\sin^2\varphi_\text{A}^{(i)}\cos^2\varphi_\text{B}^{(k-i+1)}\cdots
  \sin^2\varphi_\text{A}^{(4)}\cos^2\varphi_\text{B}^{(k-3)}
  \nonumber\\
&\quad\phantom{{}>{}}
{}\times  \Bigl(
  J_3-\cos^2\varphi_\text{A}^{(3)}\sin^2\varphi_\text{B}^{(k-3)}J_2
  \Bigr).
\label{A:irr1}
\end{align}
The last factor on the right most hand is shown to be positive definite
\begin{multline}
J_3-\cos^2\varphi_\text{A}^{(3)}\sin^2\varphi_\text{B}^{(k-3)}J_2\\
>\Bigl(
 \sin^2\varphi_\text{A}^{(3)}\cos^2\varphi_\text{B}^{(k-2)}+1
 \Bigr)
 J_2-d_3^2>0,
 \label{A:pd}
\end{multline}
which means that the quantity on the left-hand side in (\ref{A:irr1}) is positive definite.
We conclude that $J_i$ is positive definite and therefore $J_k$ does not vanish, which completes the proof.

\end{document}